\documentclass[pra,twocolumn,showpacs]{revtex4}
\usepackage{amssymb}
\usepackage{graphicx,amsmath}

\begin{document}
\date{\today}

\title{Dynamics of Bose-Einstein condensates in a one-dimensional optical lattice with double-well potential}

\author{Hanlei Zheng, Qiang Gu}
\address{Department of Physics, University of Science and Technology Beijing,
 Beijing 100083, China}

\date{\today}

\begin{abstract}
We study dynamical behaviors of the weakly interacting Bose-Einstein
condensate in the one-dimensional optical lattice with an overall
double-well potential by solving the time-dependent Gross-Pitaevskii
equation. It is observed that the double-well potential dominates
the dynamics of such a system even if the lattice depth is several
times larger than the height of the double-well potential. This
result suggests that the condensate flows without resistance in the
periodic lattice just like the case of a single particle moving in
periodic potentials. Nevertheless, the effective mass of atoms is
increased, which can be experimentally verified since it is
connected to the Josephson oscillation frequency. Moreover, the
periodic lattice enhances the nonlinearity of the double-well
condensate, making the condensate more ``self-trapped" in the
$\pi$-mode self-trapping regime.

\end{abstract}


\maketitle

\section{Introduction}

The experimental realization of confining cold atoms in optical
lattices (OL) provides great opportunities to understand quantum nature
of matter waves inside periodic potentials which is one of the
typical topics in condensed matter physics \cite{Morsch2006}. As is
well-known in solid-state physics, the perfect periodic potential
does not impede the movement of electrons, but gives rise to Bloch
band structures. Consequently, dynamics of Bloch electrons greatly
differ from free electrons in aspects such as effective mass, Bloch
oscillation and Landau-Zener tunneling \cite{Anderson}. The motion
of atoms in OL somehow resembles that of electrons in solids. Band
theory for cold atoms in periodic potentials has been developed for a
long time \cite{Minogin,Wilkens1991}. The Bloch oscillation and Landau-Zener
tunneling were observed in ultracold atoms before the achievement of
condensate \cite{BlochOsc1996,Dahan1996,Wilkinson1996}, confirming the
band structure of atoms in periodic potentials. The experiment with
Bose-Einstein condensates (BECs) in OL was first carried out by
Anderson and Kasevich \cite{Anderson1998} and it stimulated
considerable interests in this regard
\cite{Morsch2001,Burger2001,Cataliotti2001}.

In most experiments, BECs are first prepared in a magnetic trap
described as a harmonic potential. A standing wave is then created using laser beam. In order to observe the Bloch oscillation
or Landau-Zener tunneling, the overall harmonic potential needs to
be switched off \cite{Anderson1998,Morsch2001} since it destroys
translational invariance of the lattice and alters band structures
as a consequence. On the other hand, the overall harmonic potential
has some advantages in its own right. For a condensate in the
harmonic potential, it exhibits oscillation if its initial position
deviates from the bottom of the trap. It is observed that the
oscillation behaviors can be maintained in the presence of the OL,
but the oscillation frequency decreases, corresponding to an
increase in effective mass \cite{Burger2001,Cataliotti2001}. This
method can also be used to identify different dynamical regimes by varying the initial displacement of the BEC from the trap bottom.

In this paper we study dynamics of BECs inside a double-well (DW)
potential in the presence of a one-dimensional (1D) optical lattice.
An experimental situation may be realized by first loading BEC into a
double-well and then switching on the OL suddenly. A
question arises: how does the periodic lattice affect dynamics of the DW
condensate?

Before exploring the answer to this question, we briefly summarize
some already well-established features of dynamics of the DW
condensate in absence of the OL. First of all, it exhibits the
well-known dc, ac Josephson effects
\cite{Javanainen1986,Walls1996,Milburn1997,Ruostekoski1998,Leggett1998,Smerzi1997,Smerzi1999}
and the Shapiro effect \cite{Smerzi1999}, analogous to a
superconducting Josephson junction. Secondly, it displays some new
fascinating phenomena different from the superconducting Josephson
junction, such as the $\pi$-phase oscillations and macroscopic
quantum self-trapping (MQST) \cite{Smerzi1997,Smerzi1999,Giovanazzi2000}. Moreover,
a number of novel phenomena arising from quantum fluctuations are
predicted in theory, e.g., collapses and revivals of quantum
oscillation \cite{Walls1996}, coherence and decoherence
\cite{Vardi2001}, and the breaking of MQST state \cite{Kroha2009}.
After the first double-well experiment, which proved the coherence of BEC by interference
patterns \cite{Andrews1997}, Albiez {\it et al.} achieved a single
DW condensate and conducted a direct observation of the Josephson
oscillation and MQST \cite{Oberthaler2005}. The ac and dc Josephson
effects were also observed \cite{Levy2007}. So far, the static,
thermal and dynamical properties of the DW condensate have been
systematically investigated in experiments \cite{Gati2007}.

The paper is organized as follows. The theoretical model is presented
in Sect.~\ref{sect:model}. We employ the time-dependent Gross-Pitaevskii
equation (GPE) to describe the double-well condensate in the optical lattice.
Sect.~\ref{sect:results} gives a discussion of the numerical results of the
GPE. A brief summary is given in the last section.

\section{The model}
\label{sect:model}

Our study is based on the time-dependent Gross-Pitaevskii equation
. Numerical studies of the GPE have been carried out to
visualize dynamics of BECs in a one-dimensional optical lattice with
harmonic potential and to account for the main features of
experimental observations \cite{Burger2001,Cataliotti2001}.

The time-dependent GPE for BECs in a 1D optical lattice with a DW
potential is given by
\begin{eqnarray}
\imath\hbar \frac{\partial\psi(x;t)}{\partial t} = \left[
-\frac{\hbar^2}{2m}\nabla^2+V(x)+gN|\psi(x;t)|^2 \right] \psi(x;t),
\end{eqnarray}
where $\psi(x;t)$ is the macroscopic wave function at position $x$
and time $t$, $g$ represents the short-range interaction. The
potential has the form $V(x)=m\omega^{2}x^2/2 + V_{\rm
b}\mathrm{exp}(-x^2/p^2)+V_{ol}\mathrm{cos}^2(2\pi x/\lambda_{L})$,
where the first two terms describe the DW potential with $\omega$
being the trap frequency \cite{Oberthaler2005} and $V_b$ denoting
the barrier height, and the third term describes the OL which is
supposed to be created by a laser beam. $V_{ol}$ describes the
lattice depth and $\lambda_{L}/2$ denotes the lattice spacing with
$\lambda_{L}$ being the wave length of the laser beam. If the width
of the double-well is set to be $2x_0$, the number of lattice sites
that each well contains is $k=2x_0/\lambda_{L}$. Benefiting from the
characteristic length $l=\sqrt{\hbar/(m\omega)}$ and the
characteristic energy $\hbar\omega$ correspondingly, we have
$x^{\prime}=x/l$, $\tau=t\omega/2$, then the 1D GPE can be reduced to
a dimensionless one,
\begin{eqnarray}\label{1DGP}
\imath\frac{\partial \bar\psi(x^{\prime};t)}{\partial \tau} = \left[
    -\frac{\partial^2}{\partial x^{\prime 2}} + v(x^{\prime}) +
    \frac{g^{\prime}}{\pi}| \bar\psi(x^{\prime};\tau)|^2 \right]\bar\psi(x^{\prime};\tau),
\end{eqnarray}
where $v(x^{\prime})=x^{\prime 2} + 2v_{\rm
b}\mathrm{exp}(-x^{\prime 2}/p^{\prime
2})+2v_{ol}\mathrm{cos}^2(k\pi x^{\prime}/x^{\prime}_0)$, with
$v_{ol}=V_{ol}/({\hbar\omega})$, $v_b=V_b/({\hbar\omega})$,
$p^{\prime}=p/l$ and $x^{\prime}_0=x_0/l$. The dimensionless
interaction parameter $g^{\prime}=gNm/(l\hbar^2)$ where $N$ is the number of atoms.

The wave function must satisfy the condition
$\int^{\infty}_{-\infty}\mathrm{d}x^{\prime}|\bar\psi(x^{\prime};\tau)|^2=1$.
$n_L(\tau)=\int^0_{-\infty}\mathrm{d}x^{\prime}|\bar\psi(x^{\prime};\tau)|^2$ is the fraction of the number of atoms in the left well
and $n_R(\tau)=\int_0^{\infty}\mathrm{d}x^{\prime}|\bar\psi(x^{\prime};\tau)|^2$ in the right well\cite{2M}.
$\theta_L(\tau) = \mathrm{arctan}\frac{\int^0_{-\infty}\mathrm{d}x^{\prime} \mathrm{Im}
[\bar\psi(x^{\prime};\tau)]\rho(x^{\prime};\tau)} {\int^0_{-\infty}\mathrm{d}x^{\prime}
\mathrm{Re}[\bar\psi(x^{\prime};\tau)]\rho(x^{\prime};\tau)}$ and $\theta_R(\tau) =\mathrm{arctan}\frac{\int_0^{\infty}\mathrm{d}x^{\prime} \mathrm{Im}
[\bar\psi(x^{\prime};\tau)]\rho(x^{\prime};\tau)} {\int_0^{\infty}\mathrm{d}x^{\prime}
\mathrm{Re}[\bar\psi(x^{\prime};\tau)]\rho(x^{\prime};\tau)}$ are the phases in the left
well and the right well, respectively, with the density $\rho(x^{\prime};\tau)= \bar\psi^*(x^{\prime};\tau)\bar\psi(x^{\prime};\tau)$.

For the condensates in the double-well, $\phi_+(x^{\prime})$ and $\phi_-(x^{\prime})$ represent the ground state and
the first excited state wave functions. Their linear combinations are defined as the left
(right) well mode: $\psi_{L,R}(x^{\prime})=\frac{\phi_+(x^{\prime})\pm\phi_-(x^{\prime})}{2}$. They
satisfy the orthogonal condition $\int\mathrm{d}x^{\prime}\psi_L(x^{\prime})\psi_R(x^{\prime})=0$. The trial wave function for
obtaining the initial state can be chosen as the superposition of
$\psi_L(x^{\prime})$ and $\psi_R(x^{\prime})$ as in Ref. \protect\cite{2M}
$\bar\psi(x^{\prime};\tau)=\psi_L(\tau)\phi_L(x^{\prime})+\psi_R(\tau)\phi_R(x^{\prime})$,
where $\psi_{L(R)}(\tau) = \sqrt{n_{L(R)}(\tau)}\mathrm{e}^{i\theta_{L(R)}(\tau)}$. At time $\tau$, the
population imbalance and relative phase can be defined as $\Delta n=n_L-n_R$ and $\Delta\theta=\theta_L-\theta_R$, separately. A destructive technique to measure
$\Delta\theta$ is to release the BECs from the double-well potential after different evolution times to gain the interfere patterns \cite{Oberthaler2005}. Moreover, based on stimulated light scattering, a newly developed method can be used to detect $\Delta\theta$ nondestructively \cite{Saba2005}. An initially given
trial wave function at $\tau=0$ is constructed as $\bar\psi(x^{\prime};0)=\mathrm{e}^{i\Delta\theta(0)}\sqrt{n_L(0)}\psi_L(x^{\prime})+\sqrt{n_R(0)}\psi_R(x^{\prime})$.
$\Delta n(0)=n_L(0)-n_R(0)$ is the initial population imbalance and $\Delta\theta (0)$ the phase difference of condensates
in the double-well.

\begin{figure}[tb]
\includegraphics[width=0.75\linewidth]{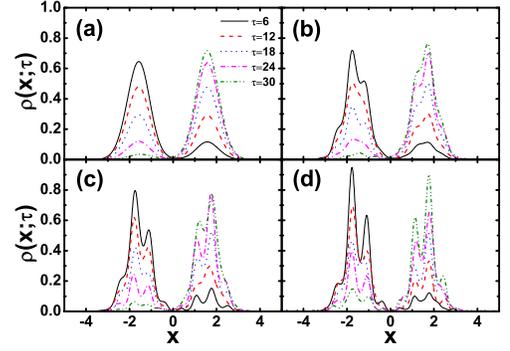}
\caption{(Color online) Snapshots of particle density distributions
in the double-well at different given times. The condensate is
initially in the Josephson oscillation state ($\Delta n=0.9$,
$\Delta\theta=0$). The height of the double-well barrier is fixed
$v_{b}=5$, but the lattice depth is tuned to be $v_{ol}=0$ (a),
$v_{ol}=5$ (b), $v_{ol}=10$ (c), $v_{ol}=15$ (d).}
\label{fig:densJO}
\end{figure}

\begin{figure}[tb]
\includegraphics[width=0.75\linewidth]{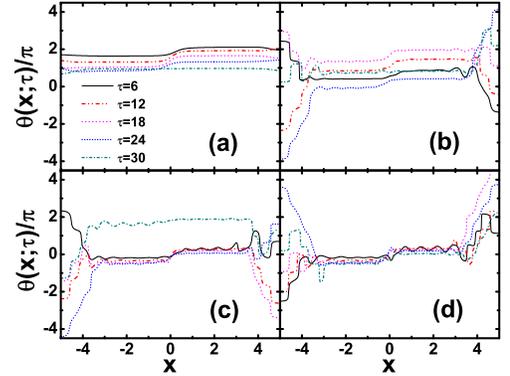}
\caption{(Color online) Snapshots of phase distributions in the
double-well at different given times. The initial state is the same
as in Fig.~\ref{fig:densJO}. $v_{b}=5$ and $v_{ol}=0$ (a),
$v_{ol}=5$ (b), $v_{ol}=10$ (c), $v_{ol}=15$ (d).}
\label{fig:phaseJO}
\end{figure}
The time-dependent GPE is numerically solved using a commonly
accepted Split-Step Crank-Nicolson discretization scheme.

\section{Results and discussions}
\label{sect:results}

In our calculation, the height of the barrier of the double-well is fixed,
$v_{b}=5$, while the lattice depth, $v_{ol}$, is tunable. One can
imagine that the double-well feature predominates the dynamics when
$v_{ol}$ is small. The phase space diagram has been systematically
calculated for the DW condensate in the absence of the OL
\cite{gu12}. At the parameters $p^{\prime}=1$ and $g^{\prime}=0.1$,
three typical dynamical regimes can be present in the phase diagram,
including Josephson oscillation, $\pi$-mode MQST and running-phase
MQST. In the following, emphasis is placed on the temporal
evolutions of two typical states; one in a Josephson oscillation
orbit around the point ($\Delta n= 0$, $\Delta \theta =0$) and one
in a $\pi$-mode MQST orbit around a point ($\Delta n>0$, $\Delta
\theta = \pi$).

Figure~\ref{fig:densJO} and \ref{fig:phaseJO} illustrate the density
and the phase distributions for the Josephson oscillation case. The
condensate is oscillating between the two wells in the absence of
the OL, as shown clearly in Fig.~\ref{fig:densJO}(a). As the optical
lattice is switched on, the density peak is split into several
narrow peaks, each of which resides in one lattice site. The narrow
peaks become more and more protrudent with increasing the lattice
depth, as shown in Fig.~\ref{fig:densJO}(b), (c) and (d). Here we
choose $k=7$ so there are $7$ sites in each well \cite{note}. Four
peaks are clearly observed in each well but the peaks near the edge
of the double-well are not high enough to be seen. Supplementary,
the lattice feature is quite obvious in the phase distribution near
the double-well edge. Without optical lattice, the phase at a given
time is almost constant in each well. In the presence of the
lattice, the phase varies slowly with position at each site, but
shows abrupt change between the neighboring sites, as shown in
Fig.~\ref{fig:phaseJO}(b), (c) and (d).

\begin{figure}[tb]
\includegraphics[width=0.75\linewidth]{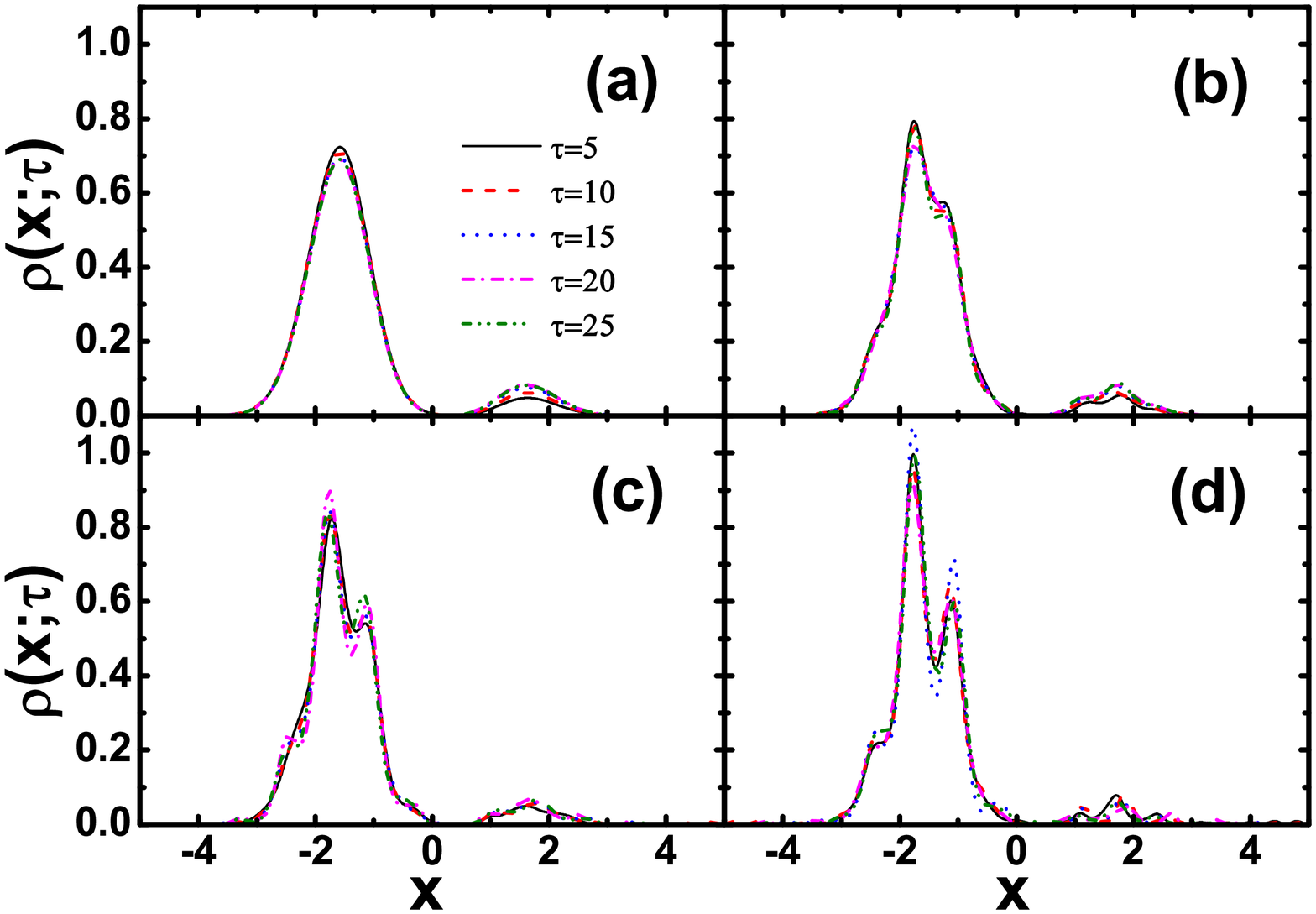}
\caption{(Color online) Snapshots of particle density distributions
in the double-well at different given times. The condensate evolves
initially from a $\pi$-mode MQST state ($\Delta n=0.9$,
$\Delta\theta=\pi$). $v_{b}=5$ and $v_{ol}=0$ (a), $v_{ol}=5$ (b),
$v_{ol}=10$ (c), $v_{ol}=15$ (d).} \label{fig:densMQST}
\label{fig:desMQST}
\end{figure}

\begin{figure}[tb]
\includegraphics[width=0.75\linewidth]{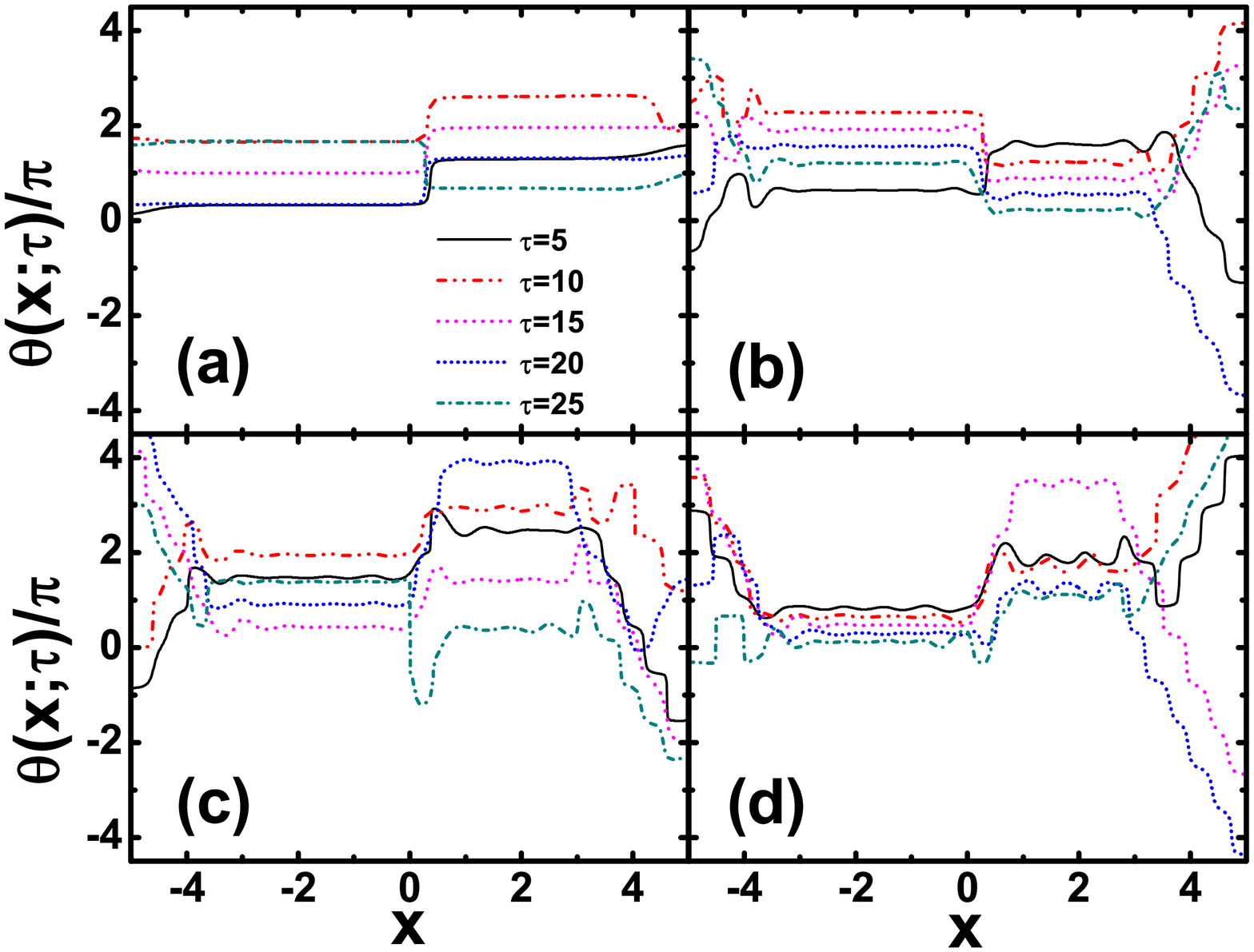}
\caption{(Color online) Snapshots of phase density distributions in
the double-well at different given times. The initial state is the
same as in Fig.~\ref{fig:desMQST}. $v_{b}=5$ and $v_{ol}=0$ (a),
$v_{ol}=5$ (b), $v_{ol}=10$ (c), $v_{ol}=15$ (d).}
\label{fig:phaseMQST}
\end{figure}

\begin{figure}[tb]
\includegraphics[width=0.75\linewidth]{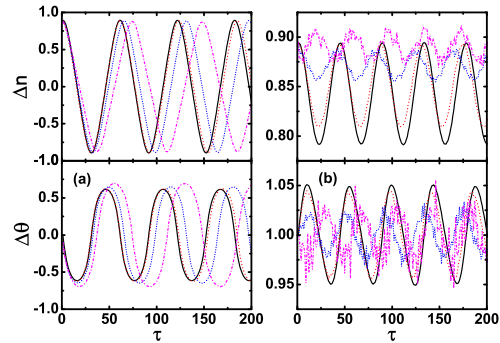}
\caption{(Color online) Evolution of $\Delta\theta$ and $\Delta n$
versus time starting from (a) a Josephson oscillation state ($\Delta
n=0.9$, $\Delta\theta=0$) and (b) a $\pi$-mode MQST state ($\Delta
n=0.9$, $\Delta\theta=\pi$). The lattice depth is $v_{ol}=0$ (solid
lines), $v_{ol}=5$ (short-dashed lines), $v_{ol}=10$ (short-dotted
lines) and $v_{ol}=15$ (short-dash-dotted lines).} \label{fig:pzt}
\end{figure}

Figures~\ref{fig:densMQST} and \ref{fig:phaseMQST} plot the
density and the phase distributions for the $\pi$-mode MQST case. Most
atoms are trapped in the left well where the density peaks are
clearly visible. On the other hand, the variation in the phase is
more pronounced in the right well. The larger the lattice depth,
the more lumpy the phase gets.

Although the optical lattice brings about remarkable changes in
density and phase distributions, the essential characters of the
double-well dynamics are not significantly affected. This phenomenon is very
surprising, because, for example, in the $v_{ol}=10$ and
$15$ cases, the DW potential is very weak comparing to the deep
OL. To illustrate this point further, we calculate temporal
evolutions of the population imbalance $\Delta n(\tau)$ and the
relative phase $\Delta\theta(\tau)$, as shown in Fig.~\ref{fig:pzt}.
For the Josephson oscillation, amplitudes of both the population
imbalance and the relative phase are just changed slightly, but their
periods have apparently been prolonged. For a pure double-well
occasion, the Josephson oscillation frequency $\omega \sim K$ where $K$ is the
tunneling constant \cite{Smerzi1997}. It is approximately inversely proportion to the mass, $m$. Therefore, the decrease of the
oscillation frequency corresponds to the increase of the effective
mass of atoms.  This result is consistent with that obtained by
studying oscillation of a BEC in the hamornic trap with the OL
\cite{Burger2001,Cataliotti2001}.

The influence of OL on the $\pi$-mode MQST seems more
notable than that on the Josephson oscillation. As shown in
Fig.~\ref{fig:pzt}(b), the evolution loses its perfect periodicity
quickly with increased lattice depth. Both $\Delta n$-$\tau$
and $\Delta\theta$-$\tau$ lines become jagged at $v_{ol}=10$ and
$15$ and the oscillation amplitudes diminish significantly. In spite
of this, the self-trapping feature remains.

\begin{figure}[tb]
\includegraphics[width=0.75\linewidth]{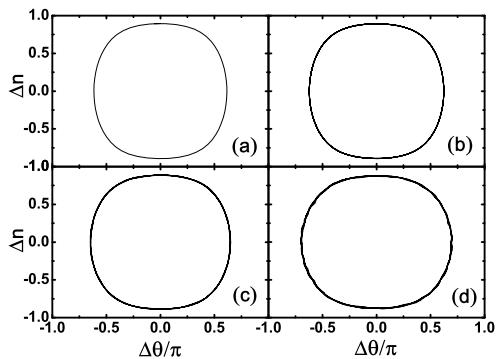}
\caption{The $\Delta\theta$-$\Delta n$ phase space diagram of a
system initially in the Josephson oscillation state ($\Delta n=0.9$,
$\Delta\theta=0$) with the frequency of optical lattice $k=7$ fixed
while optical lattice amplifications are different. (a) $v_{ol}=0$,
(b) $v_{ol}=5$, (c) $v_{ol}=10$, (d) $v_{ol}=15$.} \label{fig:JO}
\end{figure}

\begin{figure}[tb]
\includegraphics[width=0.75\linewidth]{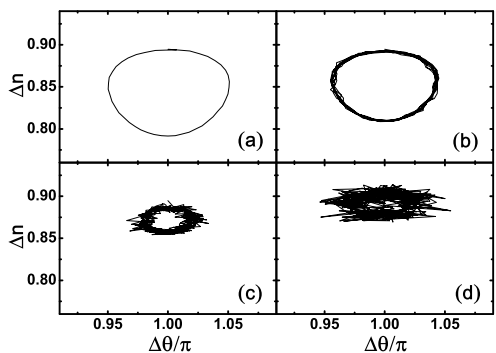}
\caption{The $\Delta\theta$-$\Delta n$ phase space diagram of a
system initially in the $\pi$-mode MQST state ($\Delta n=0.9$,
$\Delta\theta=\pi$) with the frequency of optical lattice $k=7$
fixed while optical lattice amplifications are different. (a)
$v_{ol}=0$, (b) $v_{ol}=5$, (c) $v_{ol}=10$, (d) $v_{ol}=15$.}
\label{fig:MQST}
\end{figure}

At last, we attempt to demonstrate the two typical dynamical
processes in the phase space diagram. These can be obtained by
combining $\Delta n$-$\tau$ and $\Delta\theta$-$\tau$ relations.
Fig.~\ref{fig:JO}(a) shows the phase diagram for an initial state of
Josephson oscillation ($\Delta n$=$0.9$, $\Delta\theta$=$0$), which
lies in a closed orbit circling the lowest energy point $(\Delta
n=0, \Delta\theta=0)$. With the OL imposed, the orbit circle becomes
slightly enlarged, meantime, the orbit line develops into a narrow
belt. If the lattice depth increases further, the orbit will become
fuzzy and the double-well picture will finally break down. One can
expect that the system may undergo a transition from the
superfluid state into a Mott-insulator region \cite{Jiang2007}.

For the $\pi$-mode MQST case, the orbit circle apparently shrinks when
the lattice switches on, contrary to the Josephson oscillation
case, as shown in Fig.~\ref{fig:MQST}. Meanwhile the orbit-path
becomes obscured dramatically. The ``orbit" has already lost its
definition at $v_{ol}=10$. However, the lattice compels the
condensate more ``self-trapped", and the average value of $\Delta n$
tends to grow. MQST is a nonlinear effect from the
interaction between atoms. With the interaction strength growing,
the MQST regime shrinks and moves to larger $\Delta n$ regions in the
phase space diagram \cite{Smerzi1997,gu12}. In this sense,
increasing the lattice depth yields similar results as increasing
the interaction strength. According to the tow-mode model
\cite{Smerzi1997}, the interaction strength $\Lambda=UN/2K$ grows
with the effective mass since $K$ decreases with it.

\section{Conclusion}\label{sect:conclusion}

In conclusion, we investigate the transport of BECs in a periodic
optical lattice with an overall double-well potential. We find that
the dynamics of the condensate is mainly governed by the
double-well potential even if the lattice depth is several
times larger than the height of the double-well barrier. For all
that, the periodic lattice brings about nontrivial influence. For
the Josephson oscillation state which is in the linear dynamics regime
of double-well condensates, the periodic lattice enhances
the effective atom mass which is related to the oscillation
frequency and thus the effect can be detected in experiments. For
the $\pi$-mode MQST state located in the nonlinear regimes, the
lattice effect is quite pronounced. It results in the loss of the
perfect periodicity of the $\pi$-mode oscillation and obscures the
orbit path in the phase space diagram. But, still, the
``self-trapped" feature is strengthened.

This work is supported by the National Natural Science Foundation of
China (Grant No. 11074021), and the Fundamental Research
Funds for the Central Universities of China.


\begin{thebibliography}{10}
\bibitem{note} The lattice will be more perfect if it consists of more lattice sites. We choose $k=7$
in order to show the lattice feature clearly in figures
\end{thebibliography}
\end{document}